\documentclass[aps,pra,twocolumn,showpacs]{revtex4-1}
\usepackage{graphicx,color}
\begin{document}
\title{Experimental demonstration of quantum
contextuality on an NMR qutrit}
\author{Shruti Dogra}
\email{shrutidogra@iisermohali.ac.in}
\affiliation{Department of Physical Sciences,
Indian
Institute of Science Education \&
Research (IISER) Mohali, Sector 81 SAS Nagar,
Manauli PO 140306 Punjab India.}
\author{Kavita Dorai}
\email{kavita@iisermohali.ac.in}
\affiliation{Department of Physical Sciences,
Indian
Institute of Science Education \&
Research (IISER) Mohali, Sector 81 SAS Nagar,
Manauli PO 140306 Punjab India.}
\author{Arvind}
\email{arvind@iisermohali.ac.in}
\affiliation{Department of Physical Sciences,
Indian
Institute of Science Education \&
Research (IISER) Mohali, Sector 81 SAS Nagar,
Manauli PO 140306 Punjab India.}
\begin{abstract}
We experimentally test quantum contextuality of a
single qutrit using NMR.  The contextuality
inequalities based on nine observables developed
by Kurzynski et. al. are first reformulated in
terms of traceless observables which can be
measured in an NMR experiment.  These inequalities
reveal the contextuality  of almost all
single-qutrit states.  We demonstrate the
violation of the inequality on four different
initial states of a spin-1 deuterium nucleus
oriented in a liquid crystal matrix, and follow
the violation as the states evolve in time.  We
also describe and experimentally perform a
single-shot test of contextuality for a subclass
of qutrit states whose density matrix is diagonal
in the energy basis.
\end{abstract}
\maketitle
\section{Introduction}
The existence of contextuality in the quantum description of
physical reality is a fundamental departure from any
classical theory.  For classical systems, a joint
probability distribution exists for the results of any set
of joint measurements on the system, and the results of a
measurement of a variable do not depend on the measurements
of other compatible variables.  In non-contextual hidden
variable theories, one can hence pre-assign the values to
measurement outcomes of observables before the measurement
is actually performed~\cite{peres-book}.  Quantum mechanics
on the other hand, precludes such a description of physical
reality.  In quantum mechanics there exists a context among
the measurement outcomes, which forbids us from arriving at
joint probability distributions of more than two
observables.  Given observables $A$, $B$ and $C$, such that
$A$ commutes with both $B$ and $C$, while $[B,C]\neq 0$;  a
measurement of $A$ along with $B$ and a measurement of $A$
along with $C$, may lead to different measurement outcomes
for $A$. Thus, to be able to make quantum mechanical
predictions about the outcome of a measurement, the context
of the measurement needs to be specified.

The first test for contextuality was proposed by Kochen and
Specker~\cite{kochen-jmm-1967}, wherein they used a set of
117 rays to reveal the contextuality of a single qutrit (the
KS theorem).  A modified KS scheme  based on 33 rays was
constructed by Peres~\cite{peres-jpa-1991}.  Since then,
there have been various proposals to reduce the number of
observables required to demonstrate quantum contextuality.
For state-dependent contextuality tests, only a subset of
quantum states show contextuality while other measurements
can be explained by deterministic noncontextual hidden
variable models~\cite{kcbs-prl-2008}.  State-independent
contextuality tests can be represented by an orthogonality
graph~\cite{cabello-prl-2015}.  A state-independent test of
quantum contextuality on a qutrit using a set of thirteen
projectors was developed~\cite{yu-prl-2012} and was later proved
to be optimal~\cite{cabello-pra-2012}.  
Kurzynski et.~al.~\cite{kurzynski-pra-2012} showed that
in the case of a qutrit, a contextuality
test based on a set of nine measurements is able to
reveal the contextuality of all single-qutrit states
except the maximally mixed state (which saturates
their constructed inequality).
More recent studies have
focused on testing the contextuality of indistinguishable
particles and mixed qutrit
states~\cite{su-sr-2015,xu-pra-2015}.

Experimental implementations of  contextuality
tests have been performed by different groups
using photonic
qutrits~\cite{zu-prl-2012,thompson-sr-2013,huang-pra-2013}.
KS inequalities on qubits have been tested
experimentally using a solid-state ensemble NMR
quantum computer~\cite{moussa-prl-2010}.
Hardy-like quantum contextuality has been
experimentally observed by performing sequential
measurements on photons~\cite{marques-prl-2014}.
Furthermore, the connection of contextuality with
computational speedup via magic state distillation
has been explored~\cite{howard-nature-2014}, and
the use of a single qutrit as a quantum
computational resource has also been
experimentally
demonstrated~\cite{dogra-pla-14,gedik-sr-2015}.

In the current paper, we experimentally
demonstrate the contextuality of a qutrit using
NMR.  On the basis of a set of nine measurements,
an experimental test for contextuality is designed
and implemented on an NMR qutrit.  This involves
recasting the original Kurzynski
inequality~\cite{kurzynski-pra-2012} in terms of
traceless observables, which can be measured in an
NMR experiment. The Gell-Mann matrices provide a
natural set to be used in this new scheme, as they
are  traceless and hence measurable by NMR.  We
use a deuterium nucleus (spin-1) oriented in a
liquid crystalline environment as  the NMR qutrit,
with the effective quadrupole moment of the spin
contributing to two non-overlapping resonances in
the NMR spectrum at thermal equilibrium.

The material in this paper is arranged as follows:
Section~\ref{qutrit_context} describes the
single-qutrit state-independent contextuality
inequality and the reformulation of this
inequality in terms of an experimentally feasible
set of measurements in NMR.
Section~\ref{experimental} presents the
experimental implementation of single-qutrit
contextuality, and Section~\ref{conc} contains a
few concluding remarks.
\section{State-independent test with nine observables}
\label{qutrit_context}
\begin{figure}[ht]
\centering
\includegraphics[scale=1]{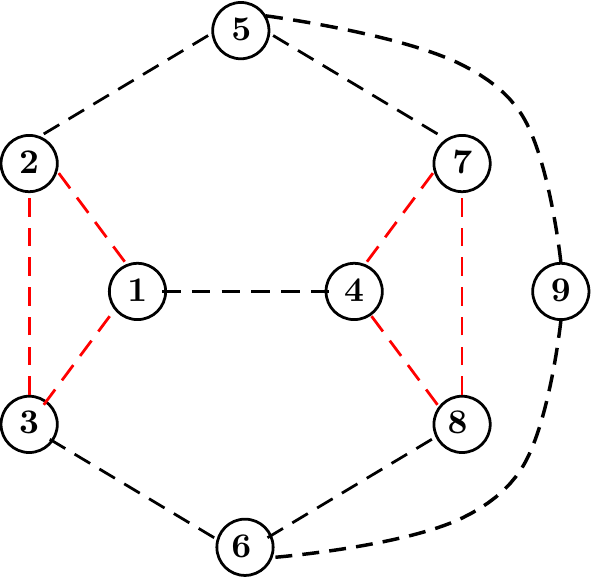}
\caption{An orthogonality graph $G$ with the 
nine projectors as the vertices and the edges $E(G)$
denoting the orthogonality relations between different
vertices. An edge between two vertices indicates
that the corresponding projectors are orthogonal.}
\label{graph}
\end{figure}
The contextuality test for a single qutrit
proposed by Kurzynski et.
al.~\cite{kurzynski-pra-2012}, consists of nine
measurements that can reveal the contextuality of
all single-qutrit states (other than the maximally
mixed state represented by the identity density
operator).  The set of nine projectors are
represented as 
\begin{equation}
\Pi_i=\vert \psi_i \rangle \langle
\psi_i \vert,\quad i=1,2\cdots 9.
\end{equation} 
The vectors $\vert \psi_i\rangle$ occupy 
vertices in the orthogonality graph $G$ and the
edges connect the vertices occupied by mutually
orthogonal vectors as shown in Fig.~\ref{graph}.

As per non-contextual hidden variable theories, one
can consider a pre-assignment of the dichotomous
measurement outcomes ($0$ or  $1$) to each of the
projection  operators $\Pi_i$. The projection
operators obey the orthogonality relations
depicted in Graph $G$, due to which no two
connected vertices in Graph $G$ can simultaneously
be assigned the value $1$.  Thus the sum total of
the maximum value of the measurement outcomes that
can be obtained from Graph $G$ is $3$.  Therefore,
repeated projective measurements of $\Pi_i$ over
an ensemble of identically prepared single qutrit
states yields the following inequality:
\begin{equation}
\sum_{i=1}^{9} \langle \Pi_i \rangle \leq 3. \label{eq1}
\end{equation}
Introducing dichotomous observables $A_i=I-2\Pi_i$
with eigen values $\pm1$ associated with
projection operators $\Pi_i$ and by using the
inequality in~(\ref{eq1}) we obtain
\begin{equation}
\quad \sum_{i=1}^{9} \langle A_i \rangle \geq 3. 
\label{eq}
\end{equation}
Measurement outcomes of $A_i$ and $A_j$ which are
connected to each other by an edge in the graph G
are mutually exclusive and can be measured
simultaneously. The edge can be thought of
as defining a context, and each $A_i$ occurs in more
than one context.  

Pre-assignment of the
measurement outcomes to each of these observables
on the basis of a non-contextual hidden variable
theory, does not consider the joint probability
distributions of all those operators which are
being co-measured.  Considering the non-contextual
pre-assignments (as before), ~(\ref{eq1}) is
reformulated in terms of the contexts, represented
as the correlations between the compatible
observables $A_i, A_j$ sharing an edge and leads
to the following inequality:
\begin{equation}
\sum_{i,j \in E(G)} \langle A_i A_j \rangle 
     +\langle A_9 \rangle \geq -4 \label{eq4}
\end{equation}
The inequalities in~(\ref{eq1}) and (\ref{eq}) represent
single-qutrit contextuality inequalities in terms
of the set of nine measurements and~(\ref{eq4})
represents the contextuality inequality based on
contexts defined in the graph G.
The violation of these inequalities  indicates
the  contextual nature of a single-qutrit state
and can be experimentally tested by performing
repeated measurements of operators $A_i$ on an
ensemble of identically prepared qutrit states.
There is no unique set of nine measurements that
test the contextual nature of every single-qutrit
state. However, corresponding to every
single-qutrit state (except the maximally mixed
state), one can always find a set of nine
projectors that reveal its
contextuality~\cite{kurzynski-pra-2012}. 
\subsection{Reformulation of contextuality
inequalities in terms of traceless operators}
\label{reformulation}
In this subsection we turn to constructing tests
to verify qutrit contextuality on an NMR quantum
information processor.  Since, in NMR we can
measure only traceless observables, we need to
reformulate the inequalities developed in
~(\ref{eq1}),(\ref{eq}) and (\ref{eq4})
in terms of traceless observables. 
A natural set of traceless
observables is provided by the  eight Gell-Mann or
$\Lambda$ matrices~\cite{gell-mann-book} and they have been used for
qutrit analysis earlier~\cite{arvind-jpa-1997}. 
In the basis spanned by
the eigen states of $S_z$, namely the states $\{\vert
+1 \rangle, \vert 0 \rangle, \vert -1 \rangle \}$
(which we will follow in the rest of this paper),
these matrices   are given by:
\begin{eqnarray} 
&&
\Lambda_{1} =\left(
\begin{array}{ccc}0 & 1 & 0 \\ 1 & 0 & 0  \\ 0 & 0
& 0 \end{array}  \right)
\Lambda_{2}
=\left( \begin{array}{ccc}0 & -\iota & 0 \\ \iota
& 0 & 0  \\ 0 & 0 & 0 \end{array}  \right)
\Lambda_{3} =\left(
\begin{array}{ccc}1 & 0 & 0 \\ 0 & -1 & 0  \\ 0 &
0 & 0 \end{array}  \right) 
\nonumber \\
&&\Lambda_{4} =\left( \begin{array}{ccc}0 & 0 & 1 \\
0 & 0 & 0  \\ 1 & 0 & 0 \end{array}  \right)
\Lambda_{5} =\left( \begin{array}{ccc}0 & 0 &
-\iota \\ 0 & 0 & 0  \\ \iota & 0 & 0 \end{array}
\right)
\Lambda_{6} =\left(
\begin{array}{ccc}0 & 0 & 0 \\ 0 & 0 & 1  \\ 0 & 1
& 0 \end{array}  \right)
\nonumber \\
&&\Lambda_{7}
=\left( \begin{array}{ccc}0 & 0 & 0 \\ 0 & 0 &
-\iota  \\ 0 & \iota & 0 \end{array}  \right)
\quad
\Lambda_{8} =
\frac{1}{\sqrt{3}}\left( \begin{array}{ccc}1 & 0 &
0 \\ 0 & 1 & 0  \\ 0 & 0 & -2 \end{array}
\right)
\label{lambdamatrix}
\end{eqnarray}

At this stage we make a choice of a particular set
of nine vectors $\vert \psi_j\rangle, j=1 \cdots
9$, occupying the  vertices of
the graph G in Figure~\ref{graph}.  
The vectors that we choose are given
in the $S_z$ basis as:
\begin{eqnarray}
&&\vert \psi_1 \rangle = \left(
\begin{array}{c}
 1 \\ 0 \\ 0
\end{array}
\right), 
\vert \psi_2 \rangle = \left(
\begin{array}{c}
 0 \\ 1 \\ 0 
\end{array}
\right), 
\vert \psi_3 \rangle = \left(
\begin{array}{c}
 0 \\ 0 \\ 1
\end{array}
\right), \nonumber \\
&&\vert \psi_4 \rangle = \left(
\begin{array}{c}
 0 \\ \sqrt{\frac{1}{2}} \\ -\sqrt{\frac{1}{2}} 
\end{array}
\right), 
\vert \psi_5 \rangle = \left(
\begin{array}{c}
 \sqrt{\frac{1}{3}} \\ 0 \\
-\sqrt{\frac{2}{3}} 
\end{array}
\right), 
\vert \psi_6 \rangle = \left(
\begin{array}{c}
\sqrt{\frac{1}{3}} \\ \sqrt{\frac{2}{3}} \\ 0 
\end{array}
\right), \nonumber \\
&&\vert \psi_7 \rangle = \left(
\begin{array}{c}
 \sqrt{\frac{1}{2}} \\ \frac{1}{2} \\ \frac{1}{2} 
\end{array}
\right), 
\vert \psi_8 \rangle = \left(
\begin{array}{c}
\sqrt{\frac{1}{2}} \\ -\frac{1}{2} \\ -\frac{1}{2}
\end{array}
\right), 
\vert \psi_9 \rangle = \left(
\begin{array}{c}
\sqrt{\frac{1}{2}} \\ -\frac{1}{2} \\ \frac{1}{2} 
\end{array}
\right)
\label{set9}
\end{eqnarray}
The observables corresponding to the nine
projection operators $A_i$ for this specific case
can be written as a combination of a set of eight
$\Lambda$ matrices and the identity matrix $I$:
\begin{eqnarray}
A_1 &=& 
-\Lambda_{3}-\frac{1}{\sqrt{3}}\Lambda_{8}+\frac{I}{3}
\nonumber  \\
A_2 &=& 
\Lambda_3-\frac{1}{\sqrt{3}}\Lambda_8+\frac{I}{3}
\nonumber  \\
A_3 &=& 
\frac{2}{\sqrt{3}}\Lambda_{8}+\frac{I}{3}
\nonumber  \\
A_4 &=&  
\frac{1}{2} \left( \Lambda_{3}
+2 \Lambda_{6}
+\frac{\Lambda_{8}}{\sqrt{3}}\right)+\frac{I}{3} 
\nonumber  \\
A_5 &=& 
\frac{1}{3} \left( -\Lambda_{3}+2 \sqrt{2}\Lambda_{4}
+\sqrt{3}\Lambda_{8} \right)+\frac{I}{3}
\nonumber  \\
A_6 &=& 
\frac{1}{3} \left( -2\sqrt{2} \Lambda_{1}+\Lambda_{3}
-\sqrt{3}\Lambda_{8} \right) +\frac{I}{3}
\nonumber  \\
A_7 &=& 
\frac{1}{4}\left( -2\sqrt{2} \Lambda_{1}
-\Lambda_{3}- 2\sqrt{2} \Lambda_{4}
-2\Lambda_{6}- \frac{\Lambda_{8}}{\sqrt{3}} \right) +\frac{I}{3}
\nonumber  \\
A_8 &=& 
\frac{1}{4} \left( 
2\sqrt{2} \Lambda_{1}-\Lambda_{3}
+2\sqrt{2} \Lambda_{4}
-2\Lambda_{6}-\frac{\Lambda_{8}}{\sqrt{3}} \right) +\frac{I}{3}
\nonumber  \\
A_9 &=& 
\frac{1}{4} \left( 
2\sqrt{2} \Lambda_{1}-\Lambda_{3}-2\sqrt{2} \Lambda_{4}
+2\Lambda_{6}-\frac{\Lambda_{8}}{\sqrt{3}} \right) 
+\frac{I}{3}  \nonumber \\
\label{neweqn5}
\end{eqnarray}
Substituting these in the inequality in 
(\ref{eq}) we obtain:
\begin{equation}
\langle-2 \sqrt{2} \Lambda_{1}-3 \Lambda_{3}
+2 \sqrt{2} \Lambda_{4}+6 \Lambda_{6}-\sqrt{3} \Lambda_{8}\rangle 
\geq 0 \label{eq5}
\end{equation}
which is an inequality written in terms of
traceless observables and obeyed by any
non-contextual single qutrit state. For a
contextual single-qutrit state, the left hand side
of (\ref{eq5}) is negative, and thus
violates the inequality.

Similarly, for  the contextuality inequality for
the correlations between  compatible observables
given in~(\ref{eq4}) we first express
all the relevant products of operators in terms of
$\Lambda$ matrices.
\begin{eqnarray}
A_1 A_2 &=&
-\frac{2}{\sqrt{3}}\Lambda_{8} -\frac{1}{3}I
\nonumber  \\
A_1A_3 &=& 
\frac{1}{\sqrt{3}} \left(\Lambda_{8}
-\sqrt{3} \Lambda_{3}\right)-\frac{1}{3}I 
\nonumber \\
A_1A_4 &=& 
\frac{1}{2} \left(-\Lambda_{3}
+2 \Lambda_{6}-\frac{\Lambda_{8}}{\sqrt{3}}\right)-\frac{1}{3}I 
\nonumber \\
A_2A_3 &=& 
\frac{1}{2} \left( 2 \Lambda_{3}
+2 \frac{\Lambda_{8}}{\sqrt{3}}\right)-\frac{1}{3}I 
\nonumber  \\
A_2A_5 &=& \frac{2}{3} \left(\Lambda_3
+ \sqrt{2} \Lambda_4\right)-\frac{1}{3}I 
\nonumber  \\
A_3A_6 &=& \frac{1}{3\sqrt{2}} \left(-\Lambda_1 
+\sqrt{2}\Lambda_3 + \sqrt{6}\Lambda_8
\right)-\frac{1}{3}I 
\nonumber  \\
A_4A_7 &=& \frac{1}{2} \left(- \sqrt{2} \Lambda_{1}
+\frac{\Lambda_{3}}{2}-\sqrt{2} \Lambda_{4}
+\Lambda_{6}+\frac{\sqrt{3}}{6} \Lambda_{8}\right)-\frac{1}{3}I
\nonumber  \\
A_4A_8 &=& \frac{1}{2} \left(\sqrt{2} \Lambda_{1}
+\frac{\Lambda_{3}}{2}+\sqrt{2} \Lambda_{4}+\Lambda_{6}
+\frac{\sqrt{3}}{6}\Lambda_{8}\right)-\frac{1}{3}I 
\nonumber   \\
A_5A_7 &=& \frac{1}{2} \left(-\sqrt{2} \Lambda_{1}
-\frac{7}{6} \Lambda_{3}+\frac{\sqrt{2}}{3}\Lambda_{4}
-\Lambda_{6}+\frac{\sqrt{3}}{2}\Lambda_{8}\right) 
-\frac{1}{3}I 
\nonumber  \\
A_5A_9 &=& \frac{1}{2} \left(\sqrt{2} \Lambda_{1}
-\frac{7}{6}\Lambda_{3}+\frac{\sqrt{2}}{3}\Lambda_{4}
+\Lambda_{6}+\frac{\sqrt{3}}{2}\Lambda_{8}\right)-\frac{1}{3}I 
\nonumber   \\
A_6A_8 &=& \frac{1}{2} \left(-\frac{\sqrt{2}}{3} \Lambda_{1}
+\frac{\Lambda_{3}}{6}+\sqrt{2} \Lambda_{4}-\Lambda_{6}
-\frac{5}{6}\sqrt{3}\Lambda_{8}\right) 
-\frac{1}{3}I 
\nonumber  \\
A_6A_9 &=& \frac{1}{2} \left(-\frac{\sqrt{2}}{3} \Lambda_{1}
+\frac{\Lambda_{3}}{6}-\sqrt{2} \Lambda_{4}+\Lambda_{6}
-\frac{5}{6} \sqrt{3}\Lambda_{8}\right) 
-\frac{1}{3}I 
\nonumber \\
A_7A_8 &=& 
\frac{1}{2} \left(-\Lambda_{3}
-2\Lambda_{6}-\frac{\Lambda_{8}}{\sqrt{3}}\right)-\frac{1}{3}I 
\label{lambda}
\end{eqnarray}
Substituting these in~(\ref{eq4}) and
rearranging the terms one obtains
\begin{equation}
 \langle -2 \sqrt{2} \Lambda_{1}-3 \Lambda_{3}
+2 \sqrt{2} \Lambda_{4}+6 \Lambda_{6}-\sqrt{3} \Lambda_{8}\rangle
\geq 0 \label{eq6}
\end{equation}
This inequality is the same as the one given in
~(\ref{eq5}) and consists of the expectation
values of the $\Lambda$ matrices that can be
determined experimentally using NMR.  A global
unitary transformation on the left hand side of
inequality~(\ref{eq6}) can be used to obtain other
equivalent inequalities. 
\section{Experimental NMR demonstration of contextuality}
\label{experimental}
\subsection{The NMR qutrit}
\begin{figure}[ht]
 \centering
 \includegraphics{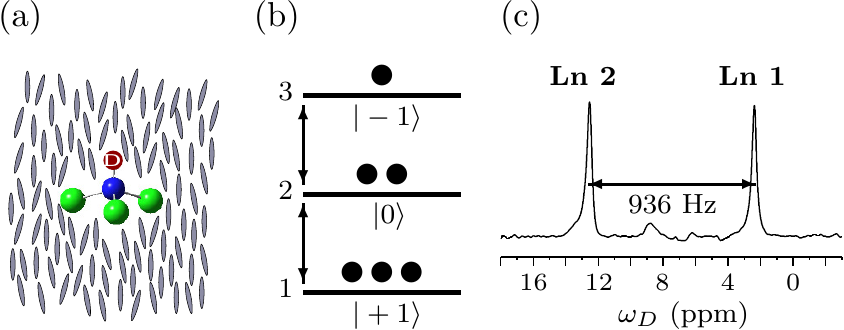}
\caption{(a) A Deuterium spin-1 nucleus oriented
in a liquid crystalline matrix.  (b) Single-qutrit
energy level diagram depicting the thermal
equilibrium population distribution of the eigen
vectors labeled as $\{1,2,3\}$ 
(corresponding to the qutrit eigenstates $\vert +1
\rangle, \vert 0 \rangle, \vert -1 \rangle$)
respectively. 
(c) Deuterium NMR
spectrum of the oriented chloroform-D molecule at
277 K. The spectral lines are labeled as {\bf Ln
1} and {\bf Ln 2}. At 277 K the NMR spectrum shows
an effective quadrupolar splitting of 936
Hz.\label{sys-con}} 
\end{figure}

The single NMR qutrit that we use for our
experiments is a spin-1 nucleus with a quadrupolar
moment, oriented in a liquid crystalline
environment.  The quadrupolar moment of this
system gets averaged to zero in an isotropically
tumbling liquid, and hence the two degenerate
single-quantum transitions in a liquid-state NMR
experiment give rise to a single resonance peak in
the NMR spectrum. This overlap of the resonances
in the NMR spectrum is undesirable for
manipulation of quantum states during a
computation.  When this spin-1 nucleus is oriented
in an anisotropic medium, the interaction of the
electric quadrupole moment of the nucleus with the
electric field gradients generated by the
surrounding electron cloud lead to an effective
quadrupolar coupling term in the
Hamiltonian~\cite{levitt-book-2008}:
\begin{equation}
H = - \omega_0 I_z
+ \lambda (3 I_z^2 - I^2) 
\end{equation}
where $\omega_0$ is the chemical shift quantifying
the Zeeman interaction and $\lambda$ 
is the effective value of the
quadrupolar coupling~\cite{yu-prl-1980}.  
The previously degenerate single-quantum transitions
now get split into two resonance peaks in the NMR
spectrum and thus can be addressed individually.
The lyotropic liquid
crystal that we use is composed of $25.6$\% of potassium
laurate, $68.16$\% of H$_2$O and $6.24$\% of
decanol; 50 $\mu$l of chloroform-D is added
to 500 $\mu$l of the liquid crystal. 
The deuterium
NMR spectrum of oriented chloroform-D was acquired
at different temperatures, as this system
possesses a liquid crystalline phase for a wide
range of temperature.  Figure~\ref{sys-con}(b) shows
the energy level diagram of the qutrit
and the relative populations in the
presence of a strong magnetic field $B_0$. The
energy levels are numbered as $\{1,2,3\}$
corresponding to the qutrit eigenstates $\vert +1
\rangle, \vert 0 \rangle, \vert -1 \rangle$
respectively, 
and the single quantum transitions
are labeled with arrows.  Figure~\ref{sys-con}(c)
shows the deuterium NMR spectrum of oriented
chloroform-D; the spectral lines corresponding to
transitions 1-2 and 2-3 are labeled {\bf Ln 1} and {\bf Ln 2}
respectively. 

The qutrit density matrix can be written as 
\begin{equation}
\rho= \frac{1}{3} I + \rho^d
\end{equation} 
In NMR experiments we have access only to the
traceless deviation density matrix
$\rho^d$, which can be expanded in 
in terms of the traceless $\Lambda$
matrices given in 
Eqn.~(\ref{lambdamatrix}) 
\begin{equation}
\rho^d = \sum_{i=1}^{8} \langle 
\Lambda_i \rangle \Lambda_i,
\end{equation}
where $\langle \Lambda_i \rangle$ is the 
expectation value of $\Lambda_i$
in the state $\rho$ (which is the same as the
expectation value of $\Lambda_i$ in the
state $\rho^d$. $\rho^d$ in turn can be
written in terms of these expectations values of 
$\Lambda$ matrices as follows
\begin{eqnarray}
\rho^d &=& \frac{1}{2} \left (\begin{array}{ccc}
\langle \Lambda_3 \rangle + 
\frac{1}{\sqrt{3}} \langle \Lambda_8 \rangle & 
\langle \Lambda_1 \rangle - \iota \langle \Lambda_2 \rangle & 
\langle \Lambda_4 \rangle - \iota \langle \Lambda_5 \rangle \\
\langle \Lambda_1 \rangle + \iota \langle \Lambda_2 \rangle & 
-\langle \Lambda_3 \rangle + \frac{1}{\sqrt{3}} \langle \Lambda_8 \rangle & 
\langle \Lambda_6 \rangle - \iota \langle \Lambda_7 \rangle \\
\langle \Lambda_4 \rangle + \iota \langle \Lambda_5 \rangle & 
\langle \Lambda_6 \rangle + \iota \langle \Lambda_7 \rangle & 
-\frac{2}{\sqrt{3}} \langle \Lambda_8 \rangle
\end{array}
\right ) \nonumber \\ \label{qut-den}
\end{eqnarray}
A single-qutrit density matrix has two
single-quantum coherences (elements
$\rho^d_{12}$  and  $\rho^d_{23}$) 
that have a direct
correspondence with the two NMR single-quantum
transitions $1-2$ and $2-3$ respectively (Figure~\ref{sys-con}).
From Eqn.~(\ref{qut-den}), we
have
\begin{eqnarray}
\rho^d_{12} &=& \frac{1}{2} 
(\langle \Lambda_1 \rangle - 
\iota \langle \Lambda_2 \rangle), \nonumber \\
\rho^d_{23} &=& \frac{1}{2} 
(\langle \Lambda_6 \rangle - 
\iota \langle \Lambda_7 \rangle)
\end{eqnarray}
and the expectation values in turn are determined
by the real and imaginary parts of the intensity
of the first line
\begin{eqnarray}
\langle \Lambda_1 \rangle_{\rho}& =& 
\langle \Lambda_1 \rangle_{\rho^d} = 
\mathfrak{Re} \{ \mathbf{I}\textrm{({\bf Ln 1})}
\}\nonumber \\
\langle \Lambda_2 \rangle_{\rho} 
&=&\langle \Lambda_2 \rangle_{\rho^d} 
= - \mathfrak{Im} \{
\mathbf{I}\textrm{({\bf Ln 1})} \}
\end{eqnarray}
Similarly, from
the second spectral line, one obtains,
\begin{eqnarray}
\langle \Lambda_6 \rangle_{\rho} &=&
\langle \Lambda_6 \rangle_{\rho^d} =
\mathfrak{Re} \{ \mathbf{I}\textrm{({\bf Ln 2})} \}
\nonumber \\
\langle \Lambda_7 \rangle_{\rho} &=&
\langle \Lambda_7 \rangle_{\rho^d} =
-\mathfrak{Im} \{ \mathbf{I}\textrm{({\bf Ln 2})} \}
\end{eqnarray}
\subsection{Experimental scheme}
A set of four NMR experiments can be used to
evaluate the inequality described in~(\ref{eq6}). The experiments use
transition-selective pulses of flip angle $\theta$
about the $j^{th}$ axis on transition $r-s$, along
with non-unitary $z$-gradient pulses to kill
unwanted coherences.  The set of experiments are
designed such that they project the directly
inaccessible expectation values of the density
matrix $\rho^d$ onto the single-quantum elements
in the final state density matrix $\rho^d_k$ of
the $k^{th}$ experiment.  The protocol for finding
the expectation values is designed to directly
measure the required expectation values and avoid
full quantum state tomography.

The experiments are delineated below:
\begin{itemize}
\item
Experiment~1~:~No operation \\ 
$\begin{array}[t]{rcl}
\langle \Lambda_1 \rangle_{\rho} &=& \mathfrak{Re} \{
\mathbf{I}\textrm{({\bf Ln 1})} \}\\
\langle \Lambda_6 \rangle_{\rho} &=&
\mathfrak{Re} \{ \mathbf{I}\textrm{({\bf Ln 2})}
\}
\end{array} $ 
\item
Experiment~2~:~$(\pi_y)^{1-2}$ \\
$\begin{array}[t]{rcl} 
\langle \Lambda_4
\rangle_{\rho} &=& \mathfrak{Re} \{
\mathbf{I}\textrm{({\bf Ln 2})} \}
\end{array}
$
\item
Experiment~3~:~Grad$_z$ - $(\frac{\pi}{2}_y)^{1-2}$ \\
$\begin{array}[b]{rcl}\langle
\Lambda_3 \rangle_{\rho} &=& \mathfrak{Re} \{
\mathbf{I}\textrm{({\bf Ln 1})} \}\end{array}$
\item 
Experiment~4~:~Grad$_z$ - $(\frac{\pi}{2}_y)^{2-3}$ \\
$\begin{array}[t]{rcl} 
\langle \Lambda_8 \rangle_{\rho} 
&=& \frac{1}{\sqrt{3}} \left[ 2
\mathfrak{Im} \{ \mathbf{I}\textrm{({\bf Ln 2})}
\}\right. 
\left.+ \langle \Lambda_3 \rangle_{\rho}

\right] 
\end{array}
$ 
\end{itemize}

\begin{figure}[ht]
\centering
\includegraphics[scale=1]{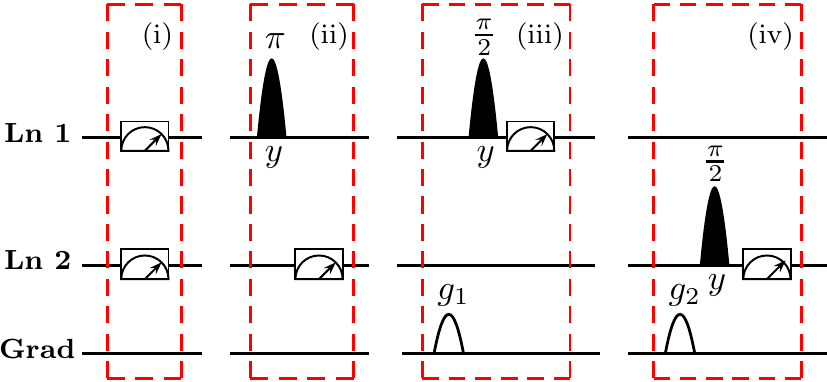}
\caption{NMR pulse sequence modules for
single-qutrit contextuality measurements.  Block
(i) yields the expectation values $\langle
\Lambda_2 \rangle$ and $\langle \Lambda_6
\rangle$, while (ii), (iii), (iv) yield the
expectation values $\langle \Lambda_4 \rangle$,
$\langle \Lambda_3 \rangle$ and  $\langle
\Lambda_8 \rangle$ respectively.  {\bf `Ln 1'} and
{\bf `Ln 2'} correspond to the two single-quantum
qutrit transitions and `Grad' denotes the gradient
channel. Pulse flip angles are shown at the top
and the axes of rotation at the bottom of each
pulse symbol~\label{pulgen}.}
\end{figure}

The expectation values obtained experimentally are
substituted into left hand side of inequality~(\ref{eq6}), to test
single-qutrit contextuality.  The set of four
experiments is shown diagrammatically in
Figure~\ref{pulgen} as four modules, numbered from
(i)-(iv).  All pulses are transition-selective
`Gaussian' shaped pulses of duration 400 $\mu s$,
applied at the frequency offsets of {\bf Ln 1}
and {\bf Ln 2} (Figure~\ref{sys-con}).  Sine-shaped 
pulsed field gradients of strength $g_i$
along the $z$ axis are applied on the gradient channel
(labeled `Grad').  The meter symbol in each
experimental module denotes a measurement of the
corresponding line intensity.  
\subsection{Testing contextuality}
\begin{figure}[ht]
\centering
\includegraphics[scale=1]{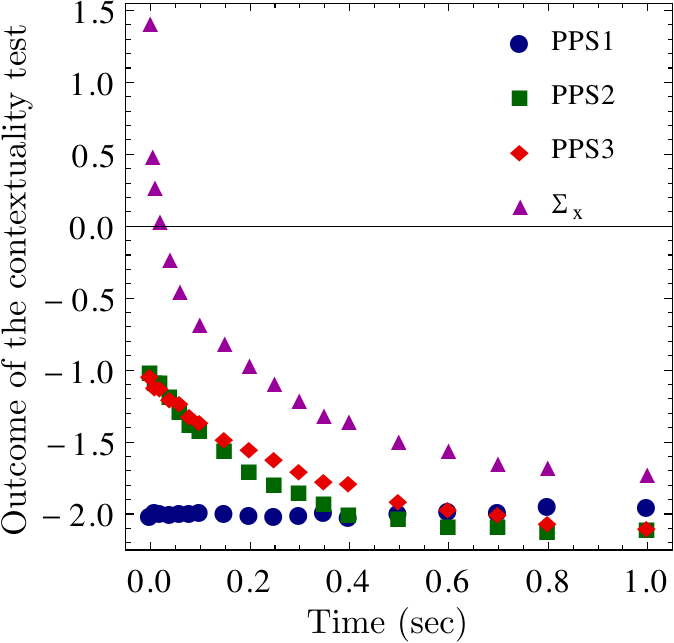}
\caption{Plot showing the  
left hand side of inequality given in~(\ref{eq6}) for four
different initial states at different time points.
The blue circles correspond to the state
`PPS1', the green square markers
markers represent the state `PPS2', the red diamonds
represent `PPS3' and the purple triangles
correspond to the state $\Sigma_x$.
\label{plot}}
\end{figure}

The experimental tests of contextuality are
performed on four different states at different
time points. The results are shown in 
Figure~\ref{plot} where the left hand side of the
non-contextuality inequality given in~(\ref{eq6}) is plotted as a
function of time. 
We describe below the four states that we have
used and their preparation schemes.
\subsubsection{PPS1}
The pseudopure state PPSI corresponding to the
pure state  $\vert +1 \rangle$ is prepared by the 
sequence:
$\left(\frac{\pi}{2}\right)_y^{(2-3)}$-Grad$_z$
applied on the thermal equilibrium state of the
single qutrit. The deviation density operator
corresponding to this state is given by 
\begin{equation}
\rho^d_{\rm PPS1} = 
\frac{1}{6}
\left(
\begin{array}{ccc}
2 & 0 & 0 \\
0 & -1 & 0 \\
0 & 0 & -1
\end{array}
\right)
\end{equation}
The left hand side of the contextuality inequality
given in~(\ref{eq6}) has the value $-2$
for this state. Therefore, the state is
contextual. The time evolution shown in
Figure~\ref{plot} reveals that 
the state moves towards the thermal equilibrium
state, 
while remaining contextual at
all times.

\subsubsection{PPS2}
The pseudopure state PPS2 corresponding to 
the pure state 
$\vert 0 \rangle$
is prepared by the
sequence:
$\left(\frac{\pi}{2}\right)_y^{(2-3)}$-Grad$_z$,
$(\pi)_y^{(1-2)}$ and the corresponding deviation
density operator is given by
\begin{equation}
\rho^d_{\rm PPS2} = 
\frac{1}{6}\left(
\begin{array}{ccc}
1 & 0 & 0 \\
0 & -2 & 0 \\
0 & 0 & 1
\end{array}
\right)
\end{equation} The left hand side of 
inequality given in~(\ref{eq6}) has the value $-1$ for this
state, showing that the PPS2 state is contextual.
As this state evolves, it tends towards thermal
equilibrium, which is reflected in the green square
markers curve in Figure~\ref{plot}.
 
\subsubsection{PPS3}
The pseudopure state PPS3 corresponding to 
the pure state 
$\vert -1  \rangle$
is prepared by the
sequence:
$\left(\frac{\pi}{2}
\right)_y^{(1-2)}$-Grad$_z$ with the corresponding
deviation density operator given by 
\begin{equation}
\rho^d_{\rm PPS3} = 
\frac{1}{6}\left (
\begin{array}{ccc}
1 & 0 & 0 \\
0 & 1 & 0 \\
0 & 0 & -2
\end{array}
\right )
\end{equation}
Inequality given in~(\ref{eq6}) has the value $-1$ for this state
showing that this state is contextual. The time
evolution takes this state towards thermal
equilibrium as in the previous cases, while keeping the
state contextual at all times.

\subsubsection{Off-Diagonal Example}
We generate a deviation density operator 
$\Sigma_x$ which does not contain
any diagonal elements, by 
a non-selective $\left(\frac{\pi}{2}\right)_y$ 
pulse on the thermal equilibrium state. 
\begin{equation}
\Sigma_x = \frac{1}{3\sqrt{2}}\left(
\begin{array}{ccc}
  0 & 1 & 0 \\
  1 & 0 & 1 \\
  0 & 1 & 0
\end{array}
\right)
\end{equation}
The contextuality value for this state is $+1.49$
and thus this state is non-contextual,
corresponding to the set of vectors in
Eqn.~(\ref{set9}).  This state approaches
thermal equilibrium, as seen from
Figure~\ref{plot}, and becomes contextual as it
evolves.
\subsection{Single-shot test for diagonal states}
The contextuality test for all the diagonal states
of a single qutrit can be performed in a 
single-shot measurement. If a state is prepared in a
diagonal form, $\langle \Lambda_1 \rangle =
\langle \Lambda_4 \rangle =\langle \Lambda_6
\rangle = 0$. Therefore for testing contextuality,
one needs to find only the non-vanishing
expectation values $\langle \Lambda_3 \rangle$ and
$\langle \Lambda_8 \rangle$. The contextuality
inequality in~(\ref{eq6}) reduces to
\begin{equation}
\langle -3 \Lambda_{3}-\sqrt{3} \Lambda_{8} \rangle \geq 0 
\label{eq6d}
\end{equation}
for this class of states.
Consider an NMR experimental test consisting of sequential
implementation of two unitary operators
$\Lambda_6(\frac{\pi}{4})$ and $\Lambda_2(\frac{\pi}{4})$ on the
diagonal state $\rho_{\rm diag}$ 
\begin{equation}
\rho_{\rm final} = e^{\iota \frac{\pi}{4} \Lambda_2} 
e^{\iota \frac{\pi}{4} \Lambda_6} \rho_{\rm diag}
e^{-\iota \frac{\pi}{4} \Lambda_6} 
e^{-\iota \frac{\pi}{4} \Lambda_2}  \label{ss}
\end{equation}
The line intensity \textbf{I}$(\mathbf{Ln 1})$
 in the NMR spectrum of state $\rho_{\rm final}$
is given by:
\begin{equation}
\mathfrak{Re}\{
\mathbf{I}(\mathbf{Ln 1})\}=
 \frac{1}{4}(-3 \langle \Lambda_{3}
\rangle_{\rho_{\rm diag}} 
-\sqrt{3} \langle\Lambda_{8}\rangle_{\rho_{\rm
diag}}) \label{int}
\end{equation}
Comparing~(\ref{eq6d}) and~(\ref{int}),
we readily obtain a simple non-contextuality
inequality for the diagonal state 
$\rho_{\rm diag} $
\begin{equation}
\mathfrak{Re} \{ \textbf{I}
\mathbf{(Ln 1)}\}_{\rm final}   \geq 0 
\label{single_shot_ineq}
\end{equation}
The violation of the above inequality is a test for 
single-qutrit contextuality, which can be carried
out in a single measurement.  

\begin{figure}
\includegraphics{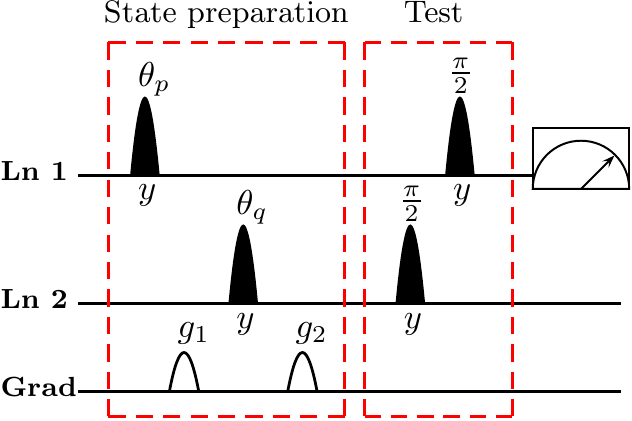}
\caption{NMR pulse sequence for the creation of an 
arbitrary single-qutrit diagonal state, followed by 
a single-shot contextuality test for diagonal  
states. {\bf `Ln 1'} and
{\bf `Ln 2'} correspond to the two single-quantum
qutrit transitions and `$Grad$' denotes the gradient
channel. Pulse flip angles are shown at the top,
and the axes of rotation at the bottom, of each
pulse symbol.}\label{pul}
\end{figure}

Various single qutrit diagonal states are obtained
from the thermal equilibrium density operator
using the pulse sequence shown in Fig.~\ref{pul}.
The set of 
deviation density matrices $\rho^d_{\rm diag}$
corresponding to the resulting diagonal states
are given by
\begin{equation}
\rho^d_{\rm diag} = 
\left(
\begin{array}{ccc}
p & 0 & 0 \\
0 & q & 0 \\
0 & 0 & r
\end{array}
\right)
\end{equation}
where $p=\cos^{2}{\frac{\theta_p}{2}}$,
$q=\cos^2{\frac{\theta_q}{2}}
\sin^2{\frac{\theta_p}{2}}-\sin^2{\frac{\theta_q}{2}}$
and $r=-(p+q)$. 
Angles $\theta_p$ and $\theta_q$ are chosen in the
range $[0, \pi]$, giving rise to arbitrary
diagonal single-qutrit states.  State preparation
is followed by the single-shot contextuality test.
The parameters used for diagonal state preparation, as well as
the theoretically expected and experimentally obtained
values for the line intensity 
$\mathfrak{Re} \{ \textbf{I}
\mathbf{(Ln 1)}\}_{\rm final}$   
in~(\ref{single_shot_ineq}) 
are shown in
Table~\ref{table1}.  
\begin{table}[h]
\caption{Parameter values of
the prepared single-qutrit diagonal states,  
values of $\theta_p$, $\theta_q$ (in
degrees).  
The theoretically expected 
$\textbf I_{\rm Th}$
and
experimentally measured $\textbf I_{\rm Exp}$
line intensity  
after the single-shot contextuality test are
also shown.
\label{table1}}
\begin{tabular}{rrrrrrrr}
\hline
No.&~~~$~~\theta_p~$& $~~~\theta_q~$ &~~~~p~
&~~~q~~& ~~~r~~& $\textbf{I}_{\rm Th}$ 
& $\textbf{I}_{\rm Exp}$\\
\hline 
1 & $0.0$ & $0.0$ & $0.33$ & $0.00$ & $-0.33$ & $-2.00$ & $-2.00$ \\
2 & $0.0$ & $90.0$ & $0.33$ & $-0.17$ & $-0.17$ & $-2.00$ & $-1.97$ \\
3 & $135.9$ & $78.4$ & $0.14$ & $0.12$ & $-0.26$ & $-0.85$ & $-0.83$ \\
4 & $140.7$ & $81.6$ & $0.11$ & $0.08$ & $-0.19$ & $-0.68$ & $-0.68$ \\
5 & $126.2$ & $76.7$ & $0.20$ & $0.11$ & $-0.31$ & $-1.23$ & $-1.13$ \\
\hline
\end{tabular} 
\end{table}
\section{Concluding Remarks}
\label{conc}
Contextuality is an inherent property of quantum
systems and has garnered much recent attention as
a computational resource. A single qutrit is the
simplest indivisible quantum system that can be
used to demonstrate quantum contextuality.  In
this work, we perform an experimental NMR test to
reveal single-qutrit contextuality, using a set of
four measurements.  The previously developed
contextuality inequality based on nine observables
was reformulated in terms of traceless observables
that can be measured by NMR.  The protocol was
successfully implemented on a single-qutrit NMR
quantum information processor  to reveal the
contextual properties of different quantum states.
A single-shot contextuality test was also devised
for the diagonal states of a qutrit.  An
experimental test of contextuality in a single
qutrit is an important step towards understanding
its strange quantum properties.
\begin{acknowledgments}
All experiments were performed on a Bruker
Avance-III 600 MHz FT-NMR spectrometer at the NMR
Research Facility at IISER Mohali.  SD
acknowledges financial support from the University
Grants Commission (UGC) India. 
\end{acknowledgments}
%

\begin{thebibliography}{22}%
\makeatletter
\providecommand \@ifxundefined [1]{%
 \@ifx{#1\undefined}
}%
\providecommand \@ifnum [1]{%
 \ifnum #1\expandafter \@firstoftwo
 \else \expandafter \@secondoftwo
 \fi
}%
\providecommand \@ifx [1]{%
 \ifx #1\expandafter \@firstoftwo
 \else \expandafter \@secondoftwo
 \fi
}%
\providecommand \natexlab [1]{#1}%
\providecommand \enquote  [1]{``#1''}%
\providecommand \bibnamefont  [1]{#1}%
\providecommand \bibfnamefont [1]{#1}%
\providecommand \citenamefont [1]{#1}%
\providecommand \href@noop [0]{\@secondoftwo}%
\providecommand \href [0]{\begingroup \@sanitize@url \@href}%
\providecommand \@href[1]{\@@startlink{#1}\@@href}%
\providecommand \@@href[1]{\endgroup#1\@@endlink}%
\providecommand \@sanitize@url [0]{\catcode `\\12\catcode `\$12\catcode
  `\&12\catcode `\#12\catcode `\^12\catcode `\_12\catcode `\%12\relax}%
\providecommand \@@startlink[1]{}%
\providecommand \@@endlink[0]{}%
\providecommand \url  [0]{\begingroup\@sanitize@url \@url }%
\providecommand \@url [1]{\endgroup\@href {#1}{\urlprefix }}%
\providecommand \urlprefix  [0]{URL }%
\providecommand \Eprint [0]{\href }%
\providecommand \doibase [0]{http://dx.doi.org/}%
\providecommand \selectlanguage [0]{\@gobble}%
\providecommand \bibinfo  [0]{\@secondoftwo}%
\providecommand \bibfield  [0]{\@secondoftwo}%
\providecommand \translation [1]{[#1]}%
\providecommand \BibitemOpen [0]{}%
\providecommand \bibitemStop [0]{}%
\providecommand \bibitemNoStop [0]{.\EOS\space}%
\providecommand \EOS [0]{\spacefactor3000\relax}%
\providecommand \BibitemShut  [1]{\csname bibitem#1\endcsname}%
\let\auto@bib@innerbib\@empty
\bibitem [{\citenamefont {Peres}(1993)}]{peres-book}%
  \BibitemOpen
  \bibfield  {author} {\bibinfo {author} {\bibfnamefont {A.}~\bibnamefont
  {Peres}},\ }\href@noop {} {\emph {\bibinfo {title} {Quantum Theory: Concepts
  and Methods}}}\ (\bibinfo  {publisher} {Kluwer Academic Publishers},\
  \bibinfo {address} {The Netherlands},\ \bibinfo {year} {1993})\BibitemShut
  {NoStop}%
\bibitem [{\citenamefont {Kochen}\ and\ \citenamefont
  {Specker}(1967)}]{kochen-jmm-1967}%
  \BibitemOpen
  \bibfield  {author} {\bibinfo {author} {\bibfnamefont {S.}~\bibnamefont
  {Kochen}}\ and\ \bibinfo {author} {\bibfnamefont {E.~P.}\ \bibnamefont
  {Specker}},\ }\href@noop {} {\bibfield  {journal} {\bibinfo  {journal} {J.
  Math. Mech.}\ }\textbf {\bibinfo {volume} {17}},\ \bibinfo {pages} {59}
  (\bibinfo {year} {1967})}\BibitemShut {NoStop}%
\bibitem [{\citenamefont {Peres}(1991)}]{peres-jpa-1991}%
  \BibitemOpen
  \bibfield  {author} {\bibinfo {author} {\bibfnamefont {A.}~\bibnamefont
  {Peres}},\ }\href@noop {} {\bibfield  {journal} {\bibinfo  {journal} {J.
  Phys. A}\ }\textbf {\bibinfo {volume} {24}},\ \bibinfo {pages} {L175}
  (\bibinfo {year} {1991})}\BibitemShut {NoStop}%
\bibitem [{\citenamefont {Klyachko}\ \emph {et~al.}(2008)\citenamefont
  {Klyachko}, \citenamefont {Can}, \citenamefont
  {Binicio\ifmmode~\breve{g}\else \u{g}\fi{}lu},\ and\ \citenamefont
  {Shumovsky}}]{kcbs-prl-2008}%
  \BibitemOpen
  \bibfield  {author} {\bibinfo {author} {\bibfnamefont {A.~A.}\ \bibnamefont
  {Klyachko}}, \bibinfo {author} {\bibfnamefont {M.~A.}\ \bibnamefont {Can}},
  \bibinfo {author} {\bibfnamefont {S.}~\bibnamefont
  {Binicio\ifmmode~\breve{g}\else \u{g}\fi{}lu}}, \ and\ \bibinfo {author}
  {\bibfnamefont {A.~S.}\ \bibnamefont {Shumovsky}},\ }\href {\doibase
  10.1103/PhysRevLett.101.020403} {\bibfield  {journal} {\bibinfo  {journal}
  {Phys. Rev. Lett.}\ }\textbf {\bibinfo {volume} {101}},\ \bibinfo {pages}
  {020403} (\bibinfo {year} {2008})}\BibitemShut {NoStop}%
\bibitem [{\citenamefont {Cabello}\ \emph {et~al.}(2015)\citenamefont
  {Cabello}, \citenamefont {Kleinmann},\ and\ \citenamefont
  {Budroni}}]{cabello-prl-2015}%
  \BibitemOpen
  \bibfield  {author} {\bibinfo {author} {\bibfnamefont {A.}~\bibnamefont
  {Cabello}}, \bibinfo {author} {\bibfnamefont {M.}~\bibnamefont {Kleinmann}},
  \ and\ \bibinfo {author} {\bibfnamefont {C.}~\bibnamefont {Budroni}},\
  }\href@noop {} {\bibfield  {journal} {\bibinfo  {journal} {Phys. Rev. Lett.}\
  }\textbf {\bibinfo {volume} {114}},\ \bibinfo {pages} {250402} (\bibinfo
  {year} {2015})}\BibitemShut {NoStop}%
\bibitem [{\citenamefont {Yu}\ and\ \citenamefont {Oh}(2012)}]{yu-prl-2012}%
  \BibitemOpen
  \bibfield  {author} {\bibinfo {author} {\bibfnamefont {S.}~\bibnamefont
  {Yu}}\ and\ \bibinfo {author} {\bibfnamefont {C.~H.}\ \bibnamefont {Oh}},\
  }\href@noop {} {\bibfield  {journal} {\bibinfo  {journal} {Phys. Rev. Lett.}\
  }\textbf {\bibinfo {volume} {108}},\ \bibinfo {pages} {030402} (\bibinfo
  {year} {2012})}\BibitemShut {NoStop}%
\bibitem [{\citenamefont {Cabello}\ \emph {et~al.}(2012)\citenamefont
  {Cabello}, \citenamefont {Amselem}, \citenamefont {Blanchfield},
  \citenamefont {Bourennane},\ and\ \citenamefont
  {Bengtsson}}]{cabello-pra-2012}%
  \BibitemOpen
  \bibfield  {author} {\bibinfo {author} {\bibfnamefont {A.}~\bibnamefont
  {Cabello}}, \bibinfo {author} {\bibfnamefont {E.}~\bibnamefont {Amselem}},
  \bibinfo {author} {\bibfnamefont {K.}~\bibnamefont {Blanchfield}}, \bibinfo
  {author} {\bibfnamefont {M.}~\bibnamefont {Bourennane}}, \ and\ \bibinfo
  {author} {\bibfnamefont {I.}~\bibnamefont {Bengtsson}},\ }\href@noop {}
  {\bibfield  {journal} {\bibinfo  {journal} {Phys. Rev. A}\ }\textbf {\bibinfo
  {volume} {85}},\ \bibinfo {pages} {032108} (\bibinfo {year}
  {2012})}\BibitemShut {NoStop}%
\bibitem [{\citenamefont {Kurzynski}\ and\ \citenamefont
  {Kaszlikowski}(2012)}]{kurzynski-pra-2012}%
  \BibitemOpen
  \bibfield  {author} {\bibinfo {author} {\bibfnamefont {P.}~\bibnamefont
  {Kurzynski}}\ and\ \bibinfo {author} {\bibfnamefont {D.}~\bibnamefont
  {Kaszlikowski}},\ }\href@noop {} {\bibfield  {journal} {\bibinfo  {journal}
  {Phys. Rev. A}\ }\textbf {\bibinfo {volume} {86}},\ \bibinfo {pages} {042125}
  (\bibinfo {year} {2012})}\BibitemShut {NoStop}%
\bibitem [{\citenamefont {Su}\ \emph {et~al.}(2015)\citenamefont {Su},
  \citenamefont {Chen},\ and\ \citenamefont {Liang}}]{su-sr-2015}%
  \BibitemOpen
  \bibfield  {author} {\bibinfo {author} {\bibfnamefont {H.-Y.}\ \bibnamefont
  {Su}}, \bibinfo {author} {\bibfnamefont {J.-L.}\ \bibnamefont {Chen}}, \ and\
  \bibinfo {author} {\bibfnamefont {Y.-C.}\ \bibnamefont {Liang}},\ }\href@noop
  {} {\bibfield  {journal} {\bibinfo  {journal} {Scientific Reports}\ }\textbf
  {\bibinfo {volume} {5}},\ \bibinfo {pages} {11637} (\bibinfo {year}
  {2015})}\BibitemShut {NoStop}%
\bibitem [{\citenamefont {Xu}\ \emph {et~al.}(2015)\citenamefont {Xu},
  \citenamefont {Su},\ and\ \citenamefont {Chen}}]{xu-pra-2015}%
  \BibitemOpen
  \bibfield  {author} {\bibinfo {author} {\bibfnamefont {Z.-P.}\ \bibnamefont
  {Xu}}, \bibinfo {author} {\bibfnamefont {H.-Y.}\ \bibnamefont {Su}}, \ and\
  \bibinfo {author} {\bibfnamefont {J.-L.}\ \bibnamefont {Chen}},\ }\href@noop
  {} {\bibfield  {journal} {\bibinfo  {journal} {Phys. Rev. A}\ }\textbf
  {\bibinfo {volume} {92}},\ \bibinfo {pages} {012104} (\bibinfo {year}
  {2015})}\BibitemShut {NoStop}%
\bibitem [{\citenamefont {Zu}\ \emph {et~al.}(2012)\citenamefont {Zu},
  \citenamefont {Wang}, \citenamefont {Deng}, \citenamefont {Chang},
  \citenamefont {Liu}, \citenamefont {Hou}, \citenamefont {Yang},\ and\
  \citenamefont {Duan}}]{zu-prl-2012}%
  \BibitemOpen
  \bibfield  {author} {\bibinfo {author} {\bibfnamefont {C.}~\bibnamefont
  {Zu}}, \bibinfo {author} {\bibfnamefont {Y.-X.}\ \bibnamefont {Wang}},
  \bibinfo {author} {\bibfnamefont {D.-L.}\ \bibnamefont {Deng}}, \bibinfo
  {author} {\bibfnamefont {X.-Y.}\ \bibnamefont {Chang}}, \bibinfo {author}
  {\bibfnamefont {K.}~\bibnamefont {Liu}}, \bibinfo {author} {\bibfnamefont
  {P.-Y.}\ \bibnamefont {Hou}}, \bibinfo {author} {\bibfnamefont {H.-X.}\
  \bibnamefont {Yang}}, \ and\ \bibinfo {author} {\bibfnamefont {L.-M.}\
  \bibnamefont {Duan}},\ }\href@noop {} {\bibfield  {journal} {\bibinfo
  {journal} {Phys. Rev. Lett.}\ }\textbf {\bibinfo {volume} {109}},\ \bibinfo
  {pages} {150401} (\bibinfo {year} {2012})}\BibitemShut {NoStop}%
\bibitem [{\citenamefont {Thompson}\ \emph {et~al.}(2013)\citenamefont
  {Thompson}, \citenamefont {Pisarczyk}, \citenamefont {Kurzynski},\ and\
  \citenamefont {Kaszlikowski}}]{thompson-sr-2013}%
  \BibitemOpen
  \bibfield  {author} {\bibinfo {author} {\bibfnamefont {J.}~\bibnamefont
  {Thompson}}, \bibinfo {author} {\bibfnamefont {R.}~\bibnamefont {Pisarczyk}},
  \bibinfo {author} {\bibfnamefont {P.}~\bibnamefont {Kurzynski}}, \ and\
  \bibinfo {author} {\bibfnamefont {D.}~\bibnamefont {Kaszlikowski}},\
  }\href@noop {} {\bibfield  {journal} {\bibinfo  {journal} {Scientific
  Reports}\ }\textbf {\bibinfo {volume} {3}},\ \bibinfo {pages} {2706}
  (\bibinfo {year} {2013})}\BibitemShut {NoStop}%
\bibitem [{\citenamefont {Huang}\ \emph {et~al.}(2013)\citenamefont {Huang},
  \citenamefont {Li}, \citenamefont {Cao}, \citenamefont {Zhang}, \citenamefont
  {Zhang}, \citenamefont {Liu}, \citenamefont {Li},\ and\ \citenamefont
  {Guo}}]{huang-pra-2013}%
  \BibitemOpen
  \bibfield  {author} {\bibinfo {author} {\bibfnamefont {Y.-F.}\ \bibnamefont
  {Huang}}, \bibinfo {author} {\bibfnamefont {M.}~\bibnamefont {Li}}, \bibinfo
  {author} {\bibfnamefont {D.-Y.}\ \bibnamefont {Cao}}, \bibinfo {author}
  {\bibfnamefont {C.}~\bibnamefont {Zhang}}, \bibinfo {author} {\bibfnamefont
  {Y.-S.}\ \bibnamefont {Zhang}}, \bibinfo {author} {\bibfnamefont {B.-H.}\
  \bibnamefont {Liu}}, \bibinfo {author} {\bibfnamefont {C.-F.}\ \bibnamefont
  {Li}}, \ and\ \bibinfo {author} {\bibfnamefont {G.-C.}\ \bibnamefont {Guo}},\
  }\href@noop {} {\bibfield  {journal} {\bibinfo  {journal} {Phys. Rev. A}\
  }\textbf {\bibinfo {volume} {87}},\ \bibinfo {pages} {052133} (\bibinfo
  {year} {2013})}\BibitemShut {NoStop}%
\bibitem [{\citenamefont {Moussa}\ \emph {et~al.}(2010)\citenamefont {Moussa},
  \citenamefont {Ryan}, \citenamefont {Cory},\ and\ \citenamefont
  {Laflamme}}]{moussa-prl-2010}%
  \BibitemOpen
  \bibfield  {author} {\bibinfo {author} {\bibfnamefont {O.}~\bibnamefont
  {Moussa}}, \bibinfo {author} {\bibfnamefont {C.~A.}\ \bibnamefont {Ryan}},
  \bibinfo {author} {\bibfnamefont {D.~G.}\ \bibnamefont {Cory}}, \ and\
  \bibinfo {author} {\bibfnamefont {R.}~\bibnamefont {Laflamme}},\ }\href@noop
  {} {\bibfield  {journal} {\bibinfo  {journal} {Phys. Rev. Lett.}\ }\textbf
  {\bibinfo {volume} {104}},\ \bibinfo {pages} {160501} (\bibinfo {year}
  {2010})}\BibitemShut {NoStop}%
\bibitem [{\citenamefont {Marques}\ \emph {et~al.}(2014)\citenamefont
  {Marques}, \citenamefont {Ahrens}, \citenamefont {Nawareg}, \citenamefont
  {Cabello},\ and\ \citenamefont {Bourennane}}]{marques-prl-2014}%
  \BibitemOpen
  \bibfield  {author} {\bibinfo {author} {\bibfnamefont {B.}~\bibnamefont
  {Marques}}, \bibinfo {author} {\bibfnamefont {J.}~\bibnamefont {Ahrens}},
  \bibinfo {author} {\bibfnamefont {M.}~\bibnamefont {Nawareg}}, \bibinfo
  {author} {\bibfnamefont {A.}~\bibnamefont {Cabello}}, \ and\ \bibinfo
  {author} {\bibfnamefont {M.}~\bibnamefont {Bourennane}},\ }\href@noop {}
  {\bibfield  {journal} {\bibinfo  {journal} {Phys. Rev. Lett.}\ }\textbf
  {\bibinfo {volume} {113}},\ \bibinfo {pages} {250403} (\bibinfo {year}
  {2014})}\BibitemShut {NoStop}%
\bibitem [{\citenamefont {Howard}\ \emph {et~al.}(2014)\citenamefont {Howard},
  \citenamefont {Wallman}, \citenamefont {Veitch},\ and\ \citenamefont
  {Emerson}}]{howard-nature-2014}%
  \BibitemOpen
  \bibfield  {author} {\bibinfo {author} {\bibfnamefont {M.}~\bibnamefont
  {Howard}}, \bibinfo {author} {\bibfnamefont {J.}~\bibnamefont {Wallman}},
  \bibinfo {author} {\bibfnamefont {V.}~\bibnamefont {Veitch}}, \ and\ \bibinfo
  {author} {\bibfnamefont {J.}~\bibnamefont {Emerson}},\ }\href@noop {}
  {\bibfield  {journal} {\bibinfo  {journal} {Nature}\ }\textbf {\bibinfo
  {volume} {510}},\ \bibinfo {pages} {351} (\bibinfo {year}
  {2014})}\BibitemShut {NoStop}%
\bibitem [{\citenamefont {Dogra}\ \emph {et~al.}(2014)\citenamefont {Dogra},
  \citenamefont {Arvind},\ and\ \citenamefont {Dorai}}]{dogra-pla-14}%
  \BibitemOpen
  \bibfield  {author} {\bibinfo {author} {\bibfnamefont {S.}~\bibnamefont
  {Dogra}}, \bibinfo {author} {\bibnamefont {Arvind}}, \ and\ \bibinfo {author}
  {\bibfnamefont {K.}~\bibnamefont {Dorai}},\ }\href@noop {} {\bibfield
  {journal} {\bibinfo  {journal} {Phys. Lett. A}\ }\textbf {\bibinfo {volume}
  {378}},\ \bibinfo {pages} {3452} (\bibinfo {year} {2014})}\BibitemShut
  {NoStop}%
\bibitem [{\citenamefont {Gedik}\ \emph {et~al.}(2015)\citenamefont {Gedik},
  \citenamefont {Silva}, \citenamefont {Cakmak}, \citenamefont {Karpat},
  \citenamefont {Vidoto}, \citenamefont {Soares-Pinto}, \citenamefont
  {deAzevedo},\ and\ \citenamefont {Fanchini}}]{gedik-sr-2015}%
  \BibitemOpen
  \bibfield  {author} {\bibinfo {author} {\bibfnamefont {Z.}~\bibnamefont
  {Gedik}}, \bibinfo {author} {\bibfnamefont {I.~A.}\ \bibnamefont {Silva}},
  \bibinfo {author} {\bibfnamefont {B.}~\bibnamefont {Cakmak}}, \bibinfo
  {author} {\bibfnamefont {G.}~\bibnamefont {Karpat}}, \bibinfo {author}
  {\bibfnamefont {E.~L.~G.}\ \bibnamefont {Vidoto}}, \bibinfo {author}
  {\bibfnamefont {D.~O.}\ \bibnamefont {Soares-Pinto}}, \bibinfo {author}
  {\bibfnamefont {E.~R.}\ \bibnamefont {deAzevedo}}, \ and\ \bibinfo {author}
  {\bibfnamefont {F.~F.}\ \bibnamefont {Fanchini}},\ }\href@noop {} {\bibfield
  {journal} {\bibinfo  {journal} {Scientific Reports}\ }\textbf {\bibinfo
  {volume} {5}},\ \bibinfo {pages} {14671} (\bibinfo {year}
  {2015})}\BibitemShut {NoStop}%
\bibitem [{\citenamefont {Gell-Mann}\ and\ \citenamefont
  {Neeman}(1964)}]{gell-mann-book}%
  \BibitemOpen
  \bibfield  {author} {\bibinfo {author} {\bibfnamefont {M.}~\bibnamefont
  {Gell-Mann}}\ and\ \bibinfo {author} {\bibfnamefont {Y.}~\bibnamefont
  {Neeman}},\ }\href@noop {} {\emph {\bibinfo {title} {The eightfold way}}}\
  (\bibinfo  {publisher} {W.A. Benjamin},\ \bibinfo {address} {Michigan},\
  \bibinfo {year} {1964})\BibitemShut {NoStop}%
\bibitem [{\citenamefont {Arvind}\ \emph {et~al.}(1997)\citenamefont {Arvind},
  \citenamefont {Mallesh},\ and\ \citenamefont {Mukunda}}]{arvind-jpa-1997}%
  \BibitemOpen
  \bibfield  {author} {\bibinfo {author} {\bibnamefont {Arvind}}, \bibinfo
  {author} {\bibfnamefont {K.~S.}\ \bibnamefont {Mallesh}}, \ and\ \bibinfo
  {author} {\bibfnamefont {N.}~\bibnamefont {Mukunda}},\ }\href@noop {}
  {\bibfield  {journal} {\bibinfo  {journal} {J. Phys. A}\ }\textbf {\bibinfo
  {volume} {30}},\ \bibinfo {pages} {2417} (\bibinfo {year}
  {1997})}\BibitemShut {NoStop}%
\bibitem [{\citenamefont {Levitt}(2008)}]{levitt-book-2008}%
  \BibitemOpen
  \bibfield  {author} {\bibinfo {author} {\bibfnamefont {M.~H.}\ \bibnamefont
  {Levitt}},\ }\href@noop {} {\emph {\bibinfo {title} {Spin dynamics:Basics of
  nuclear magnetic resonance}}}\ (\bibinfo  {publisher} {John Wiley and Sons},\
  \bibinfo {address} {Chichester England},\ \bibinfo {year} {2008})\BibitemShut
  {NoStop}%
\bibitem [{\citenamefont {Yu}\ and\ \citenamefont {Saupe}(1980)}]{yu-prl-1980}%
  \BibitemOpen
  \bibfield  {author} {\bibinfo {author} {\bibfnamefont {L.~J.}\ \bibnamefont
  {Yu}}\ and\ \bibinfo {author} {\bibfnamefont {A.}~\bibnamefont {Saupe}},\
  }\href@noop {} {\bibfield  {journal} {\bibinfo  {journal} {Phys. Rev. Lett.}\
  }\textbf {\bibinfo {volume} {45}},\ \bibinfo {pages} {1000} (\bibinfo {year}
  {1980})}\BibitemShut {NoStop}%
\end{thebibliography}
\end{document}